# Superconductivity Observed in Tantalum Polyhydride at High Pressure

X. He（何鑫）[a,1,2,3], C.L. Zhang（张昌玲）[a,1,2], Z.W. Li（李芷文）[1,2], S.J. Zhang（张思佳）[1], B. S. Min（闵保森）[1,2], J. Zhang（张俊）[1,2], K. Lu（卢可）[1,2], J.F. Zhao（赵建发）[1,2], L.C. Shi（史鲁川）[1,2], Y. Peng（彭毅）[1,2], X.C. Wang（望贤成）*[1,2], S.M. Feng（冯少敏）[1], J. Song（宋静）[1,2], L.H. Wang（王鲁红）*[4,5], V. B. Prakapenka[6], S. Chariton[6], H. Z. Liu（刘浩哲）[7], C. Q. Jin（靳常青）*[1,2,3]

[1] *Beijing National Laboratory for Condensed Matter Physics, Institute of Physics, Chinese Academy of Sciences, Beijing 100190, China*

[2] *School of Physical Sciences, University of Chinese Academy of Sciences, Beijing 100190, China*

[3] *Songshan Lake Materials Laboratory, Dongguan 523808, China*

[4] *Shanghai Advanced Research in Physical Sciences, Shanghai 201203, China*

[5] *Department of Geology, University of Illinois at Urbana Champaign, Illinois 61801, USA*

[6] *Center for Advanced Radiations Sources, University of Chicago, Chicago, Illinois 60637, USA*

[7] *Center for High Pressure Science & Technology Advanced Research, Beijing 100094, China*

We report experimental discovery of tantalum polyhydride superconductor. It was synthesized at high pressure and high temperature conditions using diamond anvil cell combined with *in-situ* high pressure laser heating techniques. The superconductivity was investigated via resistance measurements at pressures. The highest superconducting transition temperature $T_c$ was found to be ~30 K at 197 GPa in the sample that was synthesized at the same pressure with ~2000 K heating. The transitions are shifted to low temperature upon applying magnetic fields that supports the superconductivity nature. The upper critical field at zero temperature $\mu_0H_{c2}(0)$ of the superconducting phase is estimated to be ~20 T that corresponds to GL coherent length ~40 Å. Our results suggest that the superconductivity may arise from *I*-43*d* phase of TaH$_3$. It is, for the first time to our best knowledge, experimental realization of superconducting hydrides for the VB group of transitional metals.

[a] These authors contributed equally

* Corresponding authors: Wangxiancheng@iphy.ac.cn; lisaliu@illinois.edu; Jin@iphy.ac.cn

Tantalum of a group of VB elements has been widely used in jet engines and electric devices due to its high melting temperature, excellent ductility and corrosion resistance[1]. In addition, tantalum's tolerance for interstitial elements makes it as a good alloy based metal for exploring new properties and functions, for example of Ta-H system investigated for the possible hydrogen storage[2-4]. The superconductivity (SC) associated with hydrogen has been extensively studied and significant progress has been made[5-7]. Based on Bardeen–Cooper–Schrieffer (BCS) theory, the SC arising from metallic hydrogen is expected to have high superconducting temperature ($T_c$) because of its high Debye temperature. Although pure hydrogen is hard to be directly metallized by pressure[8, 9], polyhydrides provide a shortcut to realize hydrogen metallization at accessible pressure due to the hydrogen chemical pre-compression effect[10, 11]. Besides sulfur hydride, polyhydrides with clathrate-like hydrogen cage structure have been experimentally reported to have SC with $T_c$ above 200 K[12-18], such as the rare earth hydrides of $LaH_{10}$ (250-260 K at 170-200 GPa)[12-14], $YH_9$ (243-262 K at 180-201 GPa)[15, 16] as well as alkali earth hydride of $CaH_6$ (210-215 K at 160-172 GPa)[17, 18]. Many other polyhydride superconductors with moderate $T_c$ also have been found[19-23], for example, $T_c$ about 83 K at 243 GPa, 71 K at 220 GPa and 70 K at 200 GPa have been observed in $HfH_n$[21], $ZrH_n$[22] and $SnH_n$[23], respectively. Since most of the 3$d$ transition metals have local spins, which tend to present magnetic fluctuations that is negative to SC, some attentions turn to the early 5$d$ transition metals, such as Hf[21, 24] and Ta[25]. The hafnium polyhydride has been experimentally reported to exhibit SC with maximum $T_c \sim$ 83 K in the previous work[21]. Here we report the synthesis of tantalum polyhydride at high pressure of ~200 GPa and the observation of SC with the maximum $T_c$ about 30 K, which is the first superconducting hydrides experimentally realized for the VB group transitional metals.

Tantalum polyhydride was synthesized at high pressure and high temperature conditions by using diamond anvil cell (DAC) high pressure techniques. The diamond anvils with culet diameter of 50 μm beveled to 300 μm were used for the high pressure experiments. T301 stainless was used as the gasket, which was pre-pressed and drilled with a hole of 300 μm in diameter. Aluminum oxide mixed with epoxy resin acted as the insulating layer that was filled into the hole, then pre-pressed and drilled to form a high pressure chamber with 40 μm in diameter. Ammonia borane (AB) as the hydrogen source was input into the chamber, which also acted as the pressure transmitting medium. The inner electrodes were made by depositing Pt foils on the surface of the anvil culet with the thickness of 0.5 μm, which are covered by tantalum foil (99.9%) with the plane size of 20 μm *20 μm and 1 μm in thickness. After clamping the DAC, it was applied to target pressure. The pressure was determined using the Raman peak shift of diamond. The details can be referred to the ATHENA procedure reported in Ref. 26.

The high pressure heating is carried out via in-situ high pressure laser heating technique. A YAG laser in a continuous mode with 1064 nm wave length was adopted for the laser heating, and the focused spot size of laser was about 5 μm in diameter. The mixture of Ta and AB was heated at 2000 K for several minutes, where the temperature was determined by fitting the black body irradiation spectra. For the high pressure electric conductivity experiments, the sample kept with the synthezied pressure was put into a MagLab system that provides synergetic extreme environments with temperatures from 300 K to 1.5 K and a magnetic field up to 9T[27,] [28]. The Van der Pauw method was employed with an applied electric current of 1 mA.

*In-situ* high-pressure x-ray diffraction (XRD) measurements were carried out by using symmetric DACs at 13-IDD of Advanced Photon Source at the Argonne National Laboratory. The x-ray beam with wave length λ = 0.3344 Å was focused down to a spot of ~3 μm in diameter. Rhenium gasket was used and the diameter of high pressure sample is ~25 μm. For the diffraction experiments, the pressure was calibrated by the equation of state for rhenium and internal pressure marker Pt. The X ray diffraction images are converted to one dimensional diffraction data with Dioptas[29].

Tantalum superhydride samples were synthesized at the high pressure and high temperature conditions. Fig. 1(a) shows the temperature dependence of resistance $R(T)$ measured at 197 GPa for sample A (Cell 1) that was synthesized at the same pressure of 197 GPa. The resistance shows a metallic behavior in the high temperature range, and drops sharply to zero at low temperature, demonstrating a superconducting transition. The inset of Fig. 1(a) shows the enlarged view of the transition. The onset superconducting $T_c$ ~30 K can be clearly determined by the right upturn of derivative of resistance over temperature. For the sample B (Cell 2) synthesized at 181 GPa, the superconducting transitions measured at different decompression pressures are displayed in Fig. 1(b). The $T_c$ is about 25.5 K at 181 GPa, which is close to that of sample A demonstrating that the superconducting phase of samples can be repeated very well. It first rises slightly to 26 K with pressure released to 162 GPa and then goes down to 25 K at 147 GPa, showing a dome like shape of $T_c$(P). Such a trend of $T_c$ dependence on pressure was also observed in $LaH_{10}$[12] and $CaH_6$[18] superconductors, which should arise from the synergistic effect of the electron density of state near the Fermi surface and the electron-phonon coupling strength tuned by pressure.

Fig. 2(a) presents the temperature dependence of resistance measured at 197 GPa and different magnetic fields $H$ for sample A. The superconducting transition is gradually suppressed when $H$ increases. The dashed line in Fig. 2(a) marks the resistance that the value is 90% of the normal state at Tc$^{onset}$. $T_c^{90\%}$ values at different $H$ can be determined by the crosses between the dashed line and resistance curves. The critical field $H_{c2}$ versus $T_c$ is plotted in Fig. 2(b). The inset of Fig. 2(b) shows the linear fitting of the $H_{c2}(T)$ data. The slope of $dH_c/dT$ is about -1.07 T/K. Thus, by using the Werthamer-Helfand-Hohenberg (WHH) formula of $\mu_0 H_{c2}(T) = -0.69 * dH_{c2}/dT |_{Tc} * T_c$ and taking $T_c^{90\%}$ = 28.5 K, the upper critical magnetic field at zero temperature of $\mu_0 H_{c2}(0)$ can be estimated to be ~21 T. Also, $\mu_0 H_{c2}(0)$ can be estimated by the Ginzburg Landau (GL) formula of $\mu_0 H_{c2}(T) = \mu_0 H_{c2}(0)(1 - (T/T_c)^2)$. After the fitting of the $\mu_0 H_{c2}(T)$ by GL formula, as shown in Fig. 2(b), $\mu_0 H_{c2}(0)$ can be obtained to be ~20 T, which is comparable with that estimated by WHH method. From the obtained value of $\mu_0 H_{c2}(0)$, the GL coherent length $\xi$ can be estimated to be ~40 Å by the equation of $\mu_0 H_{c2}(0) = \Phi_0/2\pi\xi^2$, where $\Phi_0$ = 2.067×10$^{-15}$ Web is the magnetic flux quantum.

To further investigate the superconducting phase, the in-situ high pressure x-ray diffraction experiments were carried out. Fig. 3 shows the *in-situ* high pressure x-ray diffraction pattern collected at 195 GPa and its refinement for the sample C synthesized under the same pressure of 195 GPa (Cell 3). For clarity, table I presents the synthesis and measurements details for different samples. Most of the diffraction peaks can be indexed on the basis of a cubic lattice. For different tantalum hydrides, such as TaH$_2$, TaH$_3$ and TaH$_5$, only TaH$_3$ has been reported to be a cubic phase.

Therefore, the Rietveld refinements were performed by using the *I*-43*d* TaH$_3$ structure as the initial model, and the refinements smoothly converged to *wR* = 17.7% and $R_p$ = 12.3%, respectively. The refined parameters of a = 6.413 Å and the unit cell volume *V*=231.8 Å$^3$. The Ta atoms of *I*-43*d* TaH$_3$ are located on the 16c Wyckoff positions of (0.52624, 0.02624, 0.47376), and the schematic view of the structure is shown in the inset of Fig. 3. Although the samples for the x-ray and resistance measurements are not from the same one, considering the good repeatability of our samples, our high pressure x-ray diffraction results suggest that the synthesized superconducting tantalum polyhydride should be *I*-43*d* TaH$_3$.

In fact, tantalum dihydride of TaH$_2$ can be synthesized under pressure >5 GPa[4, 25, 30]. When increasing pressure to more than 60 GPa, the hexagonal close packed (hcp) TaH$_2$ further would react with hydrogen at room temperature to form cubic *I*-43*d* phase of TaH$_3$[25]. Beside the experimental results, TaH$_3$ was theoretically predicted to be superconducting with $T_c$ ~23 K at 80 GPa[25]. To check the SC at low pressure, we carried out the synthesis of tantalum hydride at 88 GPa at which pressure TaH$_3$ can be confirmed to be synthesized. For TaH$_2$ the hydrogen atoms are located in the octahedral (O) and tetrahedral (T) interstitial sites of hcp lattice. Since the hydrogen content is estimated to be 2.2, the over stoichiometric hydrogen suggests the O-interstice or T- interstice would accommodate more than one hydrogen atoms[4, 30]. Here for the structure model of stoichiometric *I*-43*d* TaH$_3$, it can be considered as a distorted Pm-3n (Nb$_3$Sn type) structure[25], i.e. two hydrogen atoms are located in one O-interstice (the inset of Fig. 3). It is possible that the O-interstice of *I*-43*d* TaH$_3$ can accommodate even more hydrogen atoms as seen in I*m*-3*m* phase of CaH$_6$ where four hydrogen atoms are found to be accommodated in one O-interstice[31], but the hydrogen content should be dependent on the synthesized pressure. Our results suggest that higher synthesis pressure would make the cubic lattice of TaH$_3$ accommodate more hydrogen atoms, hence in consequence, a higher $T_c$ would be favored.

In summary tantalum polyhydride was successfully synthesized and found to be SC with the maximum $T_c \sim 30$ K. The upper critical field at zero temperature $\mu_0 H_{c2}(0)$ is estimated to be ~20 T. It is suggested that the superconducting phase may arise from $I$-43$d$ TaH$_3$ phase.

**Acknowledgements**

This work was supported by the National Natural Science Foundation of China (Grant No. 11921004), the National Key R&D Program of China. (Grant Nos. 2021YFA1401800 and 2022YFA1402301), and Chinese Academy of Sciences (Grant No. XDB33010200). The in situ high pressure x-ray experiments were performed at GeoSoilEnviroCARS (the University of Chicago, Sector 13), Advanced Photon Source (APS), Argonne National Laboratory. GeoSoilEnviroCARS is supported by the National Science Foundation Earth Sciences (EAR 1634415). This research used resources of the Advanced Photon Source, a U.S. Department of Energy (DOE) Office of Science User Facility operated for the DOE Office of Science by Argonne National Laboratory (Grant No. DEAC02-06CH11357)                                        .

Table I The synthesis and measurements details for different samples

| Sample No. | Culet size | Synthesis Temperature | Synthesis Pressure | Measured technology | Measured Pressure | $T_c$ |
|---|---|---|---|---|---|---|
| A | 50 $\mu m$ | ~2000 K | 197 GPa | R-T | 197 GPa | 30 K |
| B | 50 $\mu m$ | ~2000 K | 181 GPa | R-T | 181 GPa | 25.5 K |
|  |  |  |  |  | 172 GPa | 26 K |
|  |  |  |  |  | 162 GPa | 26 K |
|  |  |  |  |  | 147 GPa | 25 K |
| C | 50 $\mu m$ | ~2000 K | 195 GPa | XRD | 195 GPa | \ |

# Figure captions

Figure 1. (a) Temperature dependence of resistance for sample A (Cell 1) measured at 197 GPa. The inset is the enlarged view of resistance curve and its temperature derivative to show the superconducting transition. (b) The superconducting transition curves for sample B (Cell 2) measured at different released pressures.

Figure 2. (a) Superconducting transition for sample A (Cell 1) measured at 197 GPa and different magnetic fields. (b) The upper critical magnetic field $\mu_0H_{c2}$(T) versus $T_c$. The red line is the fitting via GL theory. The inset shows the linear fitting for the $\mu_0H_{c2}$(T) data.

Figure 3. The x-ray diffraction pattern measured under 195 GPa and the refinement. The inset is the schematic view of $I$-43$d$ TaH$_3$ structure, showing the distorted body centered cubic structure. The green octahedron accommodates two hydrogen atoms.

Fig. 1 (a, b)

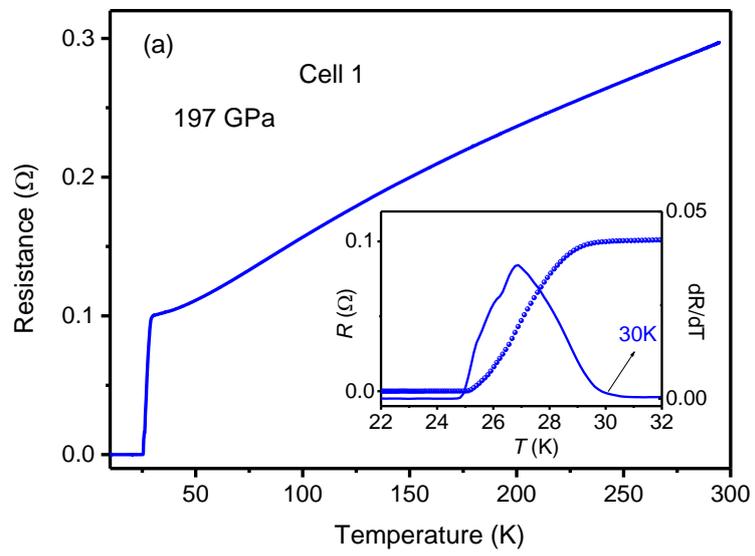

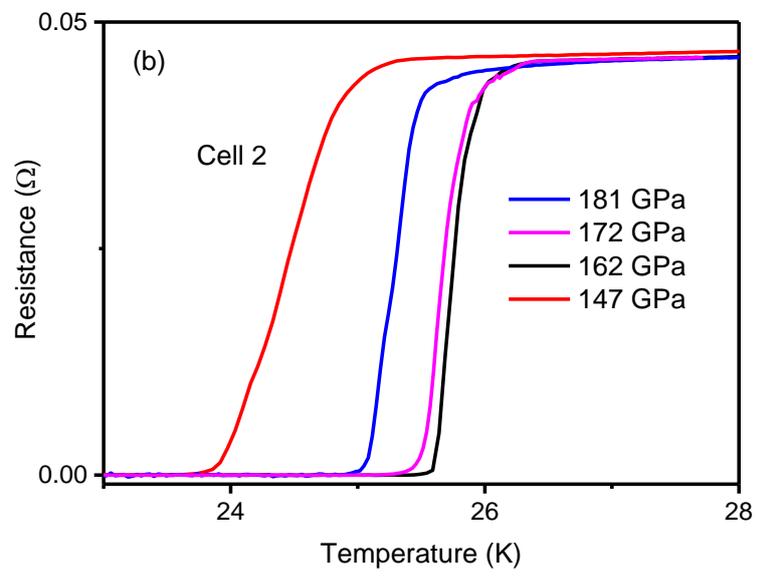

Fig. 2 (a, b)

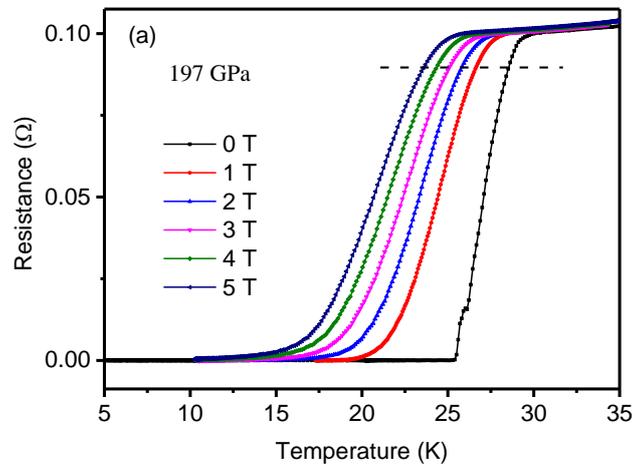

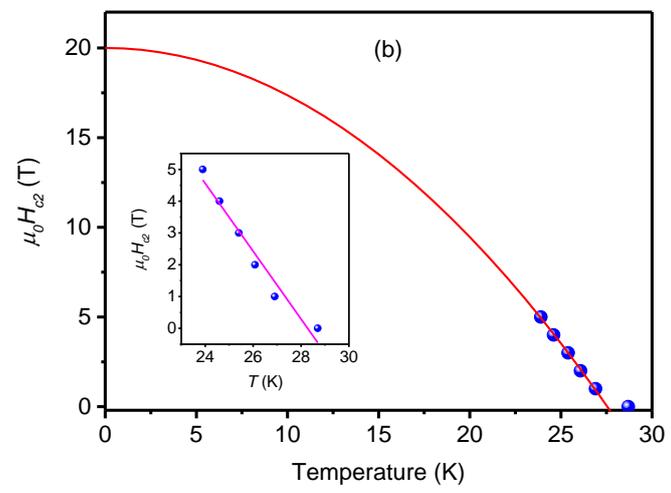

Fig. 3

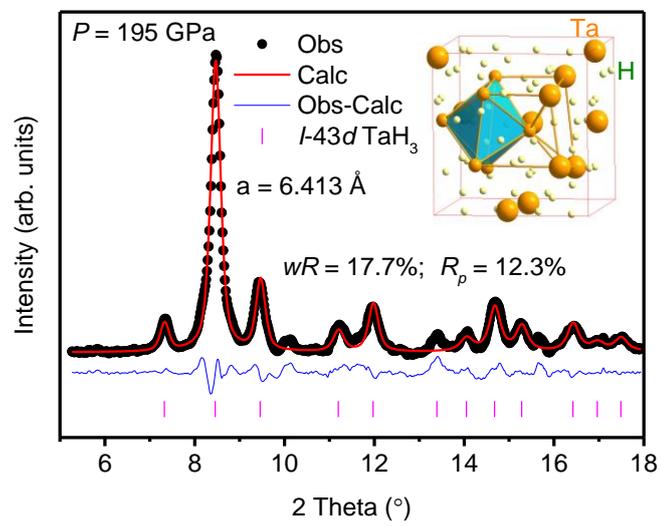